# Dramatically Accelerated Formation of Graphite Intercalation Compounds Catalyzed by Sodium

*Akira Iyo\*, Hiraku Ogino, Shigeyuki Ishida, Hiroshi Eisaki*

National Institute of Advanced Industrial Science and Technology (AIST), Tsukuba, Ibaraki 305-8568, Japan

\*Corresponding author e-mail: iyo-akira@aist.go.jp

**Abstract**

Graphite intercalation compounds (GICs) have a variety of functions due to their rich material variations, and thus, innovative methods for their synthesis are desired for practical applications. We have discovered that Na has a catalytic property that dramatically accelerates the formation of GICs. We demonstrate that $LiC_{6n}$ ($n = 1, 2$), $KC_8$, $KC_{12n}$ ($n = 2, 3, 4$), and $NaC_x$ are synthesized simply by mixing alkali metals and graphite powder with Na at room temperature (~25 °C), and $A_EC_6$ ($A_E$ = Ca, Sr, Ba) are synthesized by heating Na-added reagents at 250 °C only for a few hours. The $NaC_x$, formed by the mixing of C and Na, is understood to act as a reaction intermediate for a catalyst, thereby accelerating the formation of GICs by lowering the activation energy of intercalation. The Na-catalyzed method, which enables the rapid and mass synthesis of homogeneous GIC samples in a significantly simpler manner than conventional methods, is anticipated to stimulate research and development for GIC applications.

## 1.  Introduction

In the long history of research and development of graphite intercalation compounds (GICs), GICs have been demonstrated to include a wide variety of materials and possess numerous functions. [1-4] Based on the unique properties of GICs, their applications as highly conductive materials, [5] active materials for battery electrodes, [6] catalysts for organic synthesis, [7] gas storage [8] and superconductive materials [9] have been studied. However, only a few applications of GICs have been realized thus far, except for the use of Li-intercalated graphite in secondary batteries. [10]

For the synthesis of GICs, several methods, involving vapor-phase reaction, [2,11,12] solid-state reaction, [13] high-pressure reaction, [14] reactions in liquid reaction media, such as molten salts, [15-18] molten alloys [19,20] among others, have been developed. The vapor-phase reaction is the standard GIC synthesis method, in which GICs are synthesized by the contact of graphite with the intercalate vapor. It is the best method for the preparation of pure bulk GICs containing alkali metals (except sodium); however, it is not suitable for large-scale synthesis. During the solid-state reaction, GICs are synthesized by direct contact between the solid metal and graphite. Although relatively large GIC amounts can be synthesized using this method, the simultaneous production of undesirable carbides is also expected due to the high reaction temperature of ~500 °C. [12,14,21] In a newly developed synthesis method using LiCl-KCl molten salts, GICs are synthesized by immersing graphite sheets in the molten salts containing dissolved intercalants. [15,16] To date, bulk $BaC_6$ and $SrC_6$ have been successfully synthesized using this method. However, sample synthesis is relatively time- (up to several days) and energy-consuming. [18] Accordingly, it is desirable to develop an innovative method to efficiently synthesize large quantities of homogeneous GIC samples.

Coincidentally, while searching for new superconductors, we discovered that the formation of $CaC_6$ was significantly enhanced in samples containing a combination of Ca, C, and Na. To clarify the universality of this phenomenon, the effect of Na addition on the formation of GICs was studied for donor-type intercalants with alkali metals ($A_M$) = Li, Na, K, and alkaline earth metals ($A_E$) = Ca, Sr, Ba. Consequently, Na addition was found to be highly effective in the synthesis of $A_M$-GICs and $A_E$-GICs. Since the experimental evidence indicated that Na acts as a catalyst, the GIC synthesis method developed in this study will be referred to as the Na-catalyzed method.

This article begins with a detailed description of the $LiC_6$ synthesis process at a room temperature (RT) of ~25 °C as a typical demonstration of the effect of Na addition. Then, we propose a method for simultaneously reducing the amount of Na in the sample and fabricating sintered pellets. Next, the effectiveness of the Na-catalyzed method for the synthesis of K-GICs with controlled staging and Na-GIC itself is described. Moreover, it is demonstrated that Na acts as a potent catalyst for the synthesis of $A_E$-GIC. Finally, we discuss the catalytic function of Na and present the merits and limitations of the proposed Na-catalyzed method for GIC applications as well as future challenges.

## 2. Results And Discussion

### 2.1 Dramatic acceleration of GIC formation, Li-GICs as an example

First, the synthesis process of Li-GIC is described in detail. Li-GIC possesses a composition of $LiC_{6n}$, where $n$ (1, 2, 3, etc.) is the stage number of intercalation. Figure 1a depicts the change in the appearance of the sample that occurs as the reagents with a Li:C:Na molar ratio of 1:6:1 are mixed in a zirconia mortar using a pestle. Initially, the reagents were weakly mixed with the C powder being kneaded into the soft lumps of Na and Li. The mixing force was increased gradually. The sample then developed a metallic sheen. At this stage, the texture of the sample became soft and silvery metallic, adhering to the mortar and pestle. After continuous mixing (kneading) for ~15 min, the color of the sample changed from silvery to golden.

To investigate the reactions occurring in the mixing procedure, a portion of the sample was taken out during mixing and the XRD patterns were recorded (Figure 1b). After only ~2 min of mixing, diffraction peaks of $LiC_6$, $LiC_{12}$, and $NaC_x$ appeared, in addition to the peaks of unreacted Li, C, and Na. Figure 1c illustrates the effect of mixing time on the relative intensities of the peaks. As the mixing proceeded, Li, C, and $NaC_x$ reacted to form $LiC_6$ and $LiC_{12}$. With further mixing, the relative amount of $LiC_6$ increased, and after ~15 min of mixing, the intercalation reaction was complete, and a sample consisting only of $LiC_6$ and Na was obtained. When the reagents with the molar ratio of Li:C:Na = 1:12:2 were mixed at RT, the Li-GIC of stage 2 ($LiC_{12}$) was formed as shown below. Note that two types of stage 2 Li-GICs with different Li contents, $LiC_{12}$ and $LiC_{18}$, are known to exist. [22,23] In this study, $LiC_{12}$ was presumed to have finally formed in the sample since no Li peak was detected in the XRD pattern (Li reacted completely with C) (Figure 2c). The stage 3 Li-GIC [22,24] was not formed by mixing the reagents with Li:C:Na = 1:18:3 at RT.

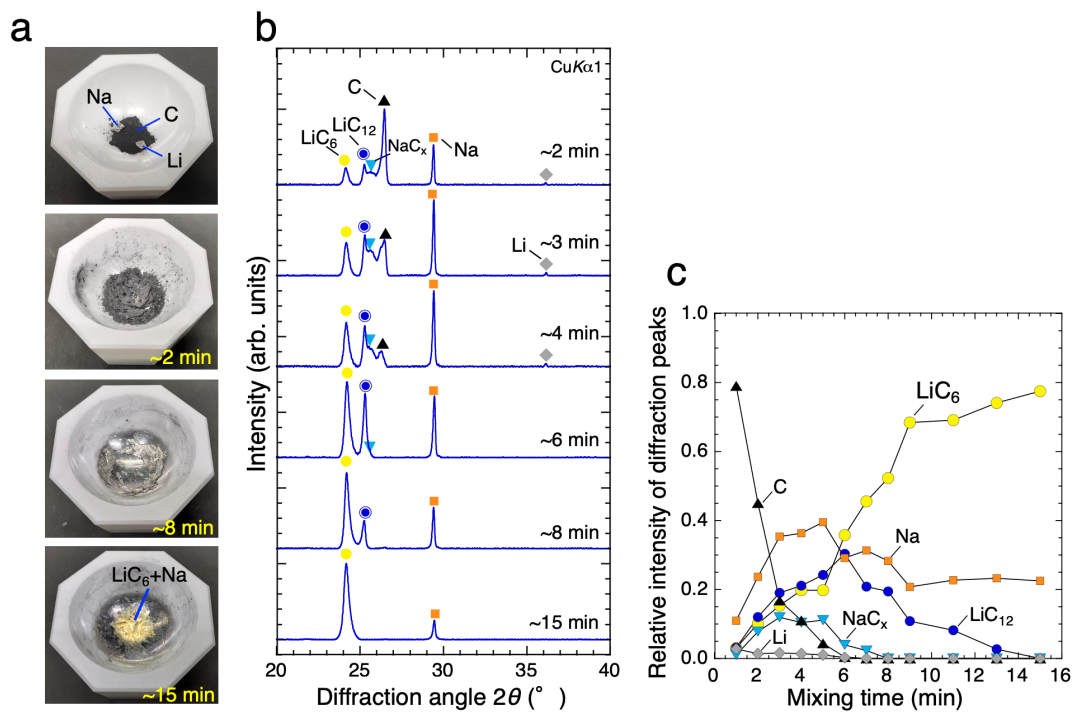

**Figure 1.** Formation process of LiC$_6$ by mixing the reagents with a molar ratio of Li:C:Na = 1:6:1 at RT. The reagents were mixed in a zirconia mortar using a pestle. **a**, Change in appearance of the sample that occurs during the mixing for 15 min. **b**, X-ray diffraction (XRD) patterns of the samples at different mixing times. **c**, Mixing time dependence of the relative intensities of the diffraction peaks for LiC$_6$, LiC$_{12}$, NaC$_x$, C, Li, and Na indicated in Figure 1b.

**2.2 Preparation of sintered GIC pellets with reduced Na content**

Na inevitably remains in the samples in the Na-catalyzed method. Herein, we propose a method for concurrently reducing Na in the sample and preparing sintered GIC pellets. Initially, the sample was placed in a tungsten carbide (WC) pellet-forming die with polytetrafluoroethylene (PTFE) sheets (NICHIAS, 0.2 mm in thickness) to prevent adhesion between the pushing rod and the sample, and weakly pressed using a pushing rod. After the die was heated to 150 °C (above the Na melting point of 98 °C) using a hot plate, it was immediately transferred to a hand press, and a pressure of ~250 MPa was applied to the pushing rod as schematically shown in Figure 2a. Since the contact between the Na-containing pellet and the PTFE sheet in the die lasted only for a few minutes, the reaction of the pellets with the PTFE sheet was weak. Nevertheless, the surface of the pellet that reacted with the PTFE sheet was removed by polishing. Approximately 80–90% of the molten Na was pushed out of the sample, and simultaneously, a homogeneous sintered pellet was obtained.

Figure 2b shows the photographs of the obtained LiC$_6$ and LiC$_{12}$ pellets, as well as graphite, for comparison. The pellets were sufficiently sintered to allow polishing of the surface. As expected, the LiC$_6$ and LiC$_{12}$ pellets were gold and dark blue in color, respectively. The XRD patterns of the powdered LiC$_6$ and LiC$_{12}$ pellets are shown in Figure

2c. Comparing the relative peak intensity of Na to LiC$_6$ in Figure 1b (bottom) and Figure 2c (top), a significant Na decrease can be seen in the sample.

The $c$-direction repeating period ($I_c$) of GICs in stage ($n$) is given by the equation:

$$I_c = d_s + (n-1)d_G \qquad (1)$$

where $d_s$ is the distance between graphite planes that sandwich the intercalant layer and $d_G$ (= 3.361 Å) is the interlayer distance of the host graphite. The parameters of the unit cell, as well as the $I_c$ and $d_s$ of GICs synthesized in this study, are summarized in Table 1. The values of LiC$_6$ and LiC$_{12}$ correspond closely to those reported previously.[14]

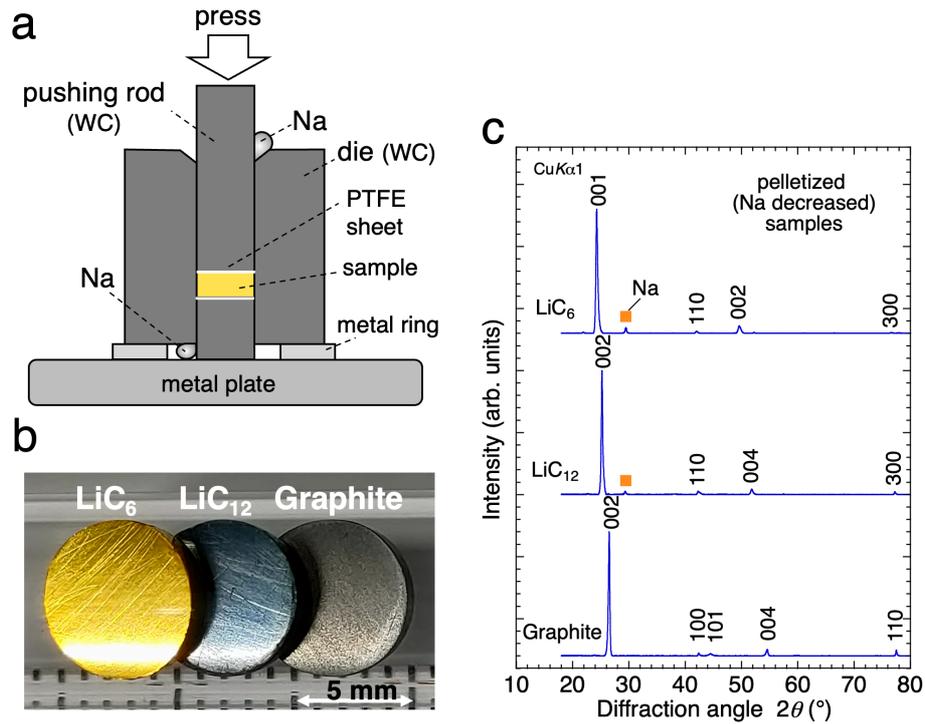

**Figure 2.** Pelletization of the sample accompanied by the reduction of Na amount in the samples by applying pressure above the melting point of Na. **a**, Schematic of pressure application to the sample using a pellet-forming die. The sample and die were heated at 150 °C prior to applying pressure. The molten Na was squeezed out of the sample through the gap between the die and the pushing rod. **b**, Photographs of LiC$_6$, LiC$_{12}$, and graphite pellets (6.7 mm in diameter and ~1.5 mm in thickness). The graphite pellet is different from that used as a reagent. **c**, Powder XRD patterns of the LiC$_6$ and LiC$_{12}$ pellets and the host graphite for comparison.

**Table 1.** Parameters of the unit cell, $I_c$, and $d_s$ of GICs synthesized in this study and the host graphite for comparison. The crystal structure models used to obtain the parameters of the unit cell are given in the references.

| GICs | $a$ (Å) | $c$ (Å) | $I_c$ (Å) | $d_s$ (Å) | Refs. |
|---|---|---|---|---|---|
| LiC$_6$ | 4.307(1) | 3.664(1) | 3.664(1) | 3.664(1) | [14] |
| LiC$_{12}$ | 4.275(1) | 7.060(3) | 7.060(3) | 3.699(3) | [14] |
| KC$_8$ | 4.963(1) | 21.426(6) | 5.357(2) | 5.357(2) | [25] |
| KC$_{24}$ | 4.938(1) | 26.240(9) | 8.747(3) | 5.386(2) | [26] |
| KC$_{36}$ | 4.933(1) | 24.116(8) | 12.058(4) | 5.336(2) | [26] |
| KC$_{48}$ | 4.930(1) | 45.930(15) | 15.310(5) | 5.227(2) | [26] |
| CaC$_6$ | 4.330(1) | 13.572(4) | 4.524(1) | 4.524(1) | [27] |
| SrC$_6$ | 4.323(1) | 9.940(3) | 4.970(2) | 4.970(2) | [28] |
| BaC$_6$ | 4.320(1) | 10.564(3) | 5.282(2) | 5.282(2) | [17] |
| Host graphite | $2a$ = 4.924(1) <br> $\sqrt{3}a$ = 4.264(1) | 6.722(3) | 3.361(1) | $d_G$ = 3.361(1) | – |

**2.3 Na-catalyzed synthesis of K-GICs and Na-GIC**

Next, the synthesis of staging-controlled K-GICs (KC$_8$ for stage 1 and KC$_{12n}$ for stage $n \geq 2$) was described. The reagent mixture (K:C:Na = 1:8:1.33) turned golden (color of KC$_8$) after only a few minutes of mixing. The fact that K and Na become liquid upon mixing at RT may have accelerated the formation of K-GICs more than that of Li-GICs. The samples were mixed for ~15 min and pelletized at 150 °C in the same manner as described for Li-GICs. Figure 3a depicts photographs of KC$_8$, KC$_{12n}$ ($n$ = 2, 3, 4), and graphite pellets. The KC$_8$ and KC$_{24}$ pellets were reddish-gold and blue, respectively. As the stage progressed to KC$_{36}$ and KC$_{48}$, the pellets became darker. Notably, KC$_{60}$ ($n$ = 5) was not formed by mixing at RT.

The corresponding XRD patterns exhibit a systematic change with progressing staging as indicated in Figure 3b. Thus, in the Na-catalyzed method, the staging of K-GICs ($n$ = 1–4), as well as Li-GICs ($n$ = 1, 2), was simply controlled by employing the same composition as the desired GICs. The parameters of the unit cell, $I_c$, and $d_s$ of K-GICs are listed in Table 1. It should be noted that the long-range stacking sequences for the high-stage ($n \geq 2$) K-GICs is not easy to determine [29], and it is not possible to determine the $c$-parameters based on the results of this study. Therefore, the $c$-parameters of K-GICs ($n$ = 2–4) in Table 1 were obtained assuming the structure models given in Ref. [26].

As $n$ increases, the $a$-parameter of the intercalated samples approaches the $a$-parameter of the host, i.e., graphite. The $d_s$ value of KC$_{12n}$ ($n$ = 2–4) decreased with increasing $n$ (the structural parameters of KC$_8$ and KC$_{12n}$ cannot be compared due to the dissimilar intercalant layer structure), which may be a result of the co-intercalation of Na, which has a smaller ionic radius than K. [30,31] However, due to residual Na in the sample (among the GIC grains), we could not use compositional analysis to confirm possible co-incarceration. Simultaneously, it has been shown that the co-intercalation of Na and Li or $A_E$ into graphite does not occur. [31,32]

It is well known that it is difficult to intercalate Na into graphite, and only high-stage NaC$_{8n}$ ($n$ = 4–8) has been synthesized thus far. [33,34] The Na-catalyzed method presented in this study was also effective in producing Na-GIC itself. Mixing the reagents (Na:C = 1:6) for ~10 min at RT yielded a silvery, soft metallic sample. A Na-GIC pellet was obtained by pressurizing the sample at 150 °C as shown in Figure 3c. The corresponding XRD patterns of the as-mixed and pelletized (Na-decreased) samples are shown in Figure 3d.

The Na-GIC pellet was black in color, and the *a*-parameter of Na-GIC (4.926 Å) was close to the corresponding value in the host graphite (2*a* = 4.924 Å). These observations suggest that the synthesized Na-GIC is a high-stage compound. However, it is difficult to determine the stage number from the XRD patterns. The primary 00*l* peaks of the Na-GICs of stages 6–8 were expected to appear at similar diffraction angles. [33,34] In addition, the Na-GIC peaks that appeared during mixing (Figure 1b and 3d) are slightly broader than the Li-GICs and K-GICs peaks. Therefore, the sample may consist of multiple stages of Na-GIC. For the aforementioned reasons, the chemical formula of Na-GIC is referred to as $NaC_x$ in this article. Note that the *d* value of the strongest $NaC_x$ peak at $2\theta$ ~25° in Figure 3d was determined to be 3.496 Å. Assuming that this peak can be ascribed to the 008 diffraction peak of $NaC_{64}$ (stage 8), [34] its $I_c$ and $d_s$ are calculated to be 27.968 Å and 4.441 Å, respectively.

Previously, K-GICs have been synthesized by mechanical mixing, in which K and graphite were stirred at 170–200 °C to produce stages *n* = 1–5. [35] We believe this is also the first report of successful $A_M$-GIC syntheses at RT by mechanical mixing.

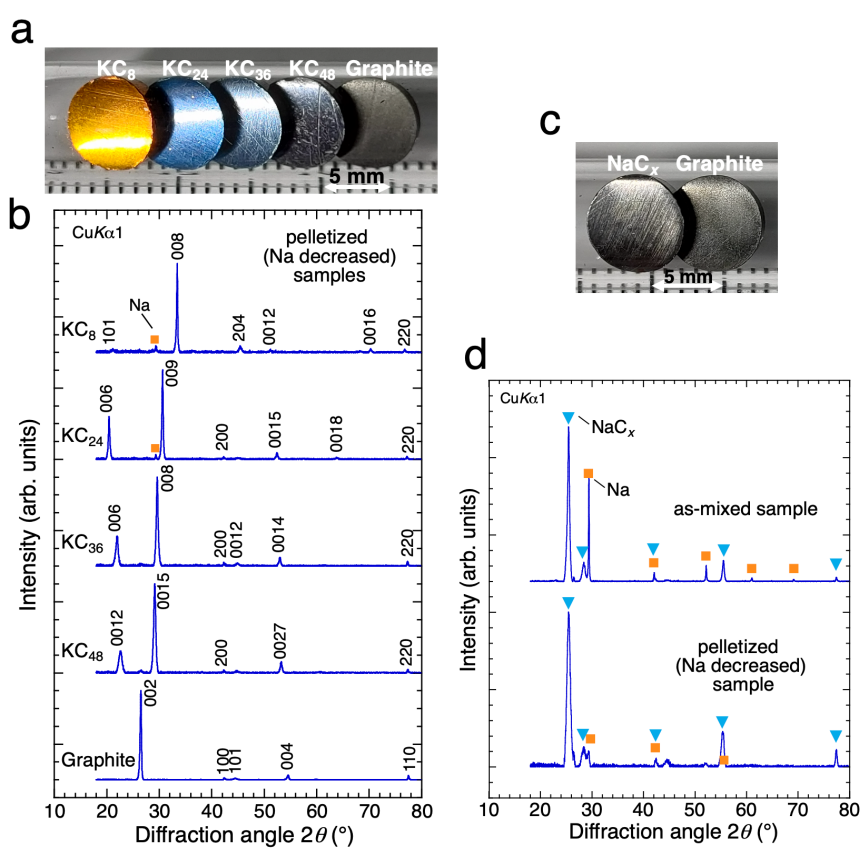

**Figure 3.** Na-catalyzed synthesis of $A_M$-GICs ($A_M$ = K, Na). **a**, Photographs of $KC_8$ and $KC_{12n}$ (*n* = 2, 3, 4) and graphite pellets. **b**, Corresponding XRD patterns of powdered $KC_8$, $KC_{12n}$ (*n* = 2, 3, 4) pellets, and host graphite. It should be noted that the XRD patterns of the pellets were almost identical to those of the as-mixed samples before pelleting, except for the decreased intensity of the Na peaks. The diffraction peaks were indexed assuming the structure given in Ref. [26] **c**, Photographs of $NaC_x$ and graphite pellets. **d**, XRD patterns of the as-mixed (Na:C = 1:6) and the pelletized (Na-decreased) $NaC_x$ samples. Note that, in contrast to Li- and K-GICs, Na-GIC in the pelletized

sample degraded due to the deintercalation of Na when it was ground in a mortar with a pestle. Accordingly, the polished surface of the pellet was subjected to the XRD measurement.

## 2.4 Remarkable effect of Na catalyst on the formation of $A_E$-GICs

$A_E$-GICs could not be formed by the RT mixing of $A_E$, C, and Na alone. For example, the as-mixed sample (Ca:C:Na = 1:6:2) included $NaC_x$, Na, and unreacted Ca as indicated in the XRD pattern in Figure 4a (top). To promote the reaction between $A_E$ and $NaC_x$, a heat treatment process was added. After mixing the reagents ($A_E$:C:Na = 1:6:2) for ~15 min, the samples were placed in the die as a reaction vessel, pressed weakly with the pushing rod, and subsequently heated at 250 °C for 2 h on a hot plate (PTFE sheets were not used in this step). Figure 4a (bottom) shows the XRD pattern of the heat-treated sample for $A_E$ = Ca. The change in the XRD patterns clearly indicates that $CaC_6$ was formed by the reaction of Ca and $NaC_x$. The unreacted Ca and $NaC_x$ can be reduced by additional heat treatment (250 °C for 2 h) with an intermediate grinding (the sample was once removed from the die, ground (kneaded) in a mortar for ~10 min, and then returned to the die). In the case of $A_E$ = Sr and Ba, the reactions were almost complete by a single heat treatment at 250 °C for only 2 h.

The heat-treated samples were pelletized in the same manner as previously shown in Figure 2a. A photograph of the obtained $A_EC_6$ ($A_E$ = Ca, Sr, Ba) pellets is shown in Figure 4b. All $A_EC_6$ pellets had a lighter golden hue than the $LiC_6$ pellets. However, $A_EC_6$ is reported to exhibit a silver shine. [17,28,36] We believe that difference in $A_E$ occupancy in $A_EC_6$ may be responsible for the color difference. As indicated in the powder XRD patterns (Figure 4c), most of the diffraction peaks of $A_EC_6$ were indexed based on the previously reported crystal structures. [27,37,38] The parameters of the unit cell, $I_c$, and $d_s$ of $A_EC_6$ ($A_E$ = Ca, Sr, Ba) listed in Table 1 are also consistent with those previously reported. [13,17,28,39,40] As shown in the inset of Figure 4c, $CaC_6$ exhibits bulk superconductivity at ~11 K (−262.2 °C), which is in line with previously reported results. [41]

Considering that $A_EC_6$ has been typically synthesized at higher temperatures (350–450 °C) and for a longer duration (6–12 days) using molten alloy and molten salt methods, [18,19,28] the proposed Na-catalyzed method significantly lowered the temperature and shortened the duration of the $A_EC_6$ synthesis. We are convinced that $NaC_x$, which is formed by mixing Na and C-containing reagents at RT, plays a crucial role in the accelerated formation of GICs. Theoretical calculations have demonstrated that $NaC_x$ is energetically unstable. [42-44] Consequently, it can be inferred that $A_M$ or $A_E$ can be rapidly intercalated into graphite layers opened by the $NaC_x$ formation, resulting in the generation of more stable $A_M$- or $A_E$-GICs. Na is eventually extruded from the graphite interlayers. Hence, Na acts as a catalyst, and $NaC_x$, as a reaction intermediate of the catalyst, decreases the activation energy for the formation of $A_M$- and $A_E$-GICs.

Na has been utilized in the synthesis of Li-GIC in the past. Basu et al., [21] and Billaud et al., [24] synthesized Li-GICs in stages 2–3 by immersing graphite in a molten Na-Li alloy at 150 °C. It was assumed that Na acts as an inert solvent that inhibits Li activity. [21] To the best of our knowledge, this is the first time an active function of Na was identified and presented as an efficient and universal method for $A_M$- and $A_E$-GIC synthesis.

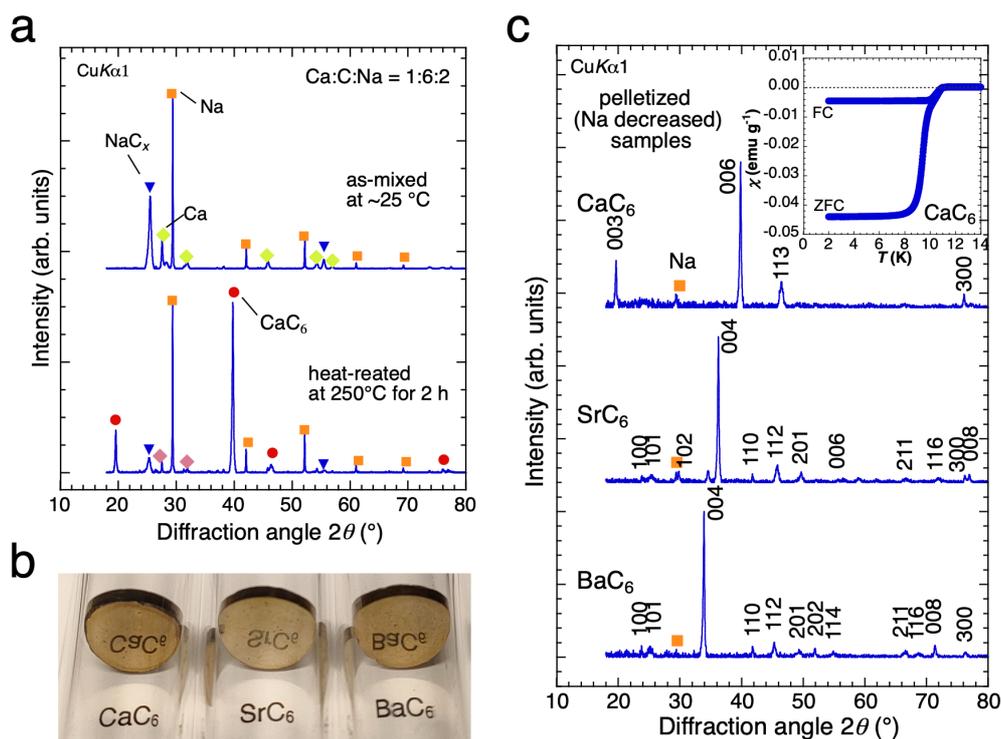

**Figure 4.** Na-catalyzed synthesis of $A_E$-GICs ($A_E$ = Ca, Sr, Ba). **a**, XRD patterns of the as-mixed and heat-treated samples (Ca:C:Na = 1:6:2). **b**, Photograph of $A_E C_6$ ($A_E$ = Ca, Sr, Ba) pellets with a diameter of 6.7 mm and thickness of ~1.5 mm. The pellet surfaces are polished reflecting the letters written on the paper. **c**, XRD patterns of $A_E C_6$ ($A_E$ = Ca, Sr, Ba) pellets. The inset shows magnetic susceptibility $\chi$ of the $CaC_6$ pellet.

## 3. Conclusion

We have demonstrated that the catalytic effect of Na dramatically accelerates the formation of $A_M$-GICs and $A_E$-GICs. The $NaC_x$ appears to act as a reaction intermediate, lowering the activation energy for the GIC formation. Due to the following features, the Na-catalyzed method is promising for the mass synthesis of homogeneous GIC samples for application. First, the graphite powder that was used as a host material is less expensive than highly oriented graphite, which is typically used for GIC synthesis. Moreover, samples can be synthesized by mixing at RT or heat treatments at a low temperature (250 °C), which reduces the number of facilities required for sample preparation. Furthermore, the time required for sample synthesis is drastically reduced from days to hours in comparison to conventional methods. Since the research on Na-catalyzed GIC synthesis has only begun, there are still numerous issues to be investigated. For instance, it would be desirable to have a deeper understanding of how $A_M$- and $A_E$-GICs are formed from the high-stage $NaC_x$. Additionally, it would be of great interest to apply the Na-catalyzed method to synthesize GICs using other intercalators, such as heavy alkali metals and lanthanides. Finally, the catalytic function of Na itself is of interest for future research, as it may be effective in improving the performance of secondary batteries, which use the graphite intercalation phenomenon.

## 4. Experimental Section

**Materials.** Graphite (Furuuchi chemical, 99.99%, -200 mesh, Lot. 80124) was utilized as the intercalation host. Prior to sample synthesis, we confirmed that the use of the graphite with and without heat treatment did not affect the final results; therefore, the heat treatment was not performed to simplify the synthesis process. For the intercalants $A_M$ and $A_E$, pure bulk reagents (purity of 99–99.9%) were used. $A_E$ were powdered prior to weighing. Since the reagents and GICs are sensitive to air, the sample synthesis was conducted in a glovebox filled with argon gas ($O_2$ concentration less than 1 ppm, the dew point of $H_2O$ less than −70 °C).

$A_M$ or $A_E$ and C were weighed in accordance with the composition ratios of the desired GICs. The molar ratios of Na to C in $A_M$- and $A_E$-GIC were set at 1:6 and 1:3, respectively. For example, the reagents were weighed in the molar ratio of Li:C:Na = 1:6:1, Li:C:Na = 1:12:2, and Ca:C:Na = 1:6:2 for the synthesis of $LiC_6$, $LiC_{12}$, and $CaC_6$, respectively. For Li- and K-GIC, the amount of Na was chosen to sufficiently complete the formation of Li- and K-GIC by mixing at RT for 15 min. For $A_E$-GIC, twice as much Na was added since the reactivity of $A_E$ with graphite is lower than that of Li and K. The amount of Na should be adjusted according to the synthesis conditions, such as mixing time and reaction temperature. For the synthesis of Na-GIC, reagents were weighed in the molar ratio of Na:C = 1:6. Note that $NaC_{64}$ was not formed by mixing reagents with the molar ratio of Na:C = 1:64. Therefore, the Na:C ratio was set at 1:6 similar to Li-GIC and K-GIC synthesis. The weighed reagents (typically 0.1–0.2 g in total weight) were mixed and kneaded for 10–15 min at RT using a mortar and pestle. $A_M$-GICs were produced by mixing the reagents at RT. In contrast, $A_E$-GICs were additionally heated at 250 °C for a total duration of several hours after being mixed at RT. Consequently, samples consisting of $A_M$- or $A_E$-GICs and Na were obtained.

As a control, reagents were mixed (and heated) without Na at RT for 15 min. As a result, the samples mixed with only Li or K and C at RT contained multiple stages or dominated by unreacted C (Figures S1 and S2, Supporting Information). For $A_E$-GIC, $A_EC_6$ was not formed through the mixing at RT, and only trace amounts of $A_EC_6$ were formed after the heat treatment at 250 °C (Figure S3, Supporting Information). In addition, we tested whether K could also act as a catalyst in the same way as Na. However, only stable K-GICs were formed, and neither Li-GICs nor $A_E$-GICs could be formed (Figure S4, Supporting Information).

**Measurements.** The samples were evaluated using X-ray diffraction (XRD) with Cu$K\alpha$ radiation (Rigaku Ultima IV). Samples containing large amounts of Na showed clay-like behavior. In contrast, samples with a low amount of Na could be ground into powders. Both samples were mounted by pressing them strongly against the sample holders, resulting in highly oriented X-ray diffraction patterns. The XRD measurements were performed using an airtight attachment to avoid samples exposure to air, which limited the measurement range of $2\theta \geq 18°$. The background and Cu$K\alpha_2$ components of diffraction were eliminated from the diffraction data by software processing. The $a$- and $c$-parameters of GICs were determined from the $d$ values of the diffraction peaks. The magnetic susceptibility of the sample was measured using a magnetic property measurement system (Quantum Design, MPMS-XL7). Measurements were performed in a magnetic field of 10 Oe with zero-field-cooled (ZFC) and field-cooled (FC) modes.

## Acknowledgments

We are grateful to Dr. Yasumoto Tanaka for helpful discussions.

## Competing Interests

The authors declare no competing interests.

## Data Availability Statement

The data that support the findings of this study are available from the corresponding author upon reasonable request.